\documentclass[preprint2]{aastex}

\usepackage{amssymb,amsmath}

\shorttitle{Generic Structure of Magnetic Reconnection}
\shortauthors{Dumin \& Somov}

\begin{document}

\title{What is Generic Structure of the 3D Null-Point \\
       Magnetic Reconnection?}

\author{Yurii V. Dumin}
\affil{P.K.~Sternberg Astronomical Institute (GAISh)
       of M.V.~Lomonosov Moscow State University, \\
       Universitetskii prosp.\ 13, 119992, Moscow, Russia \\
and \\
       Space Research Institute (IKI)
       of Russian Academy of Sciences, \\
       Profsoyuznaya str.\ 84/32, 117997, Moscow, Russia}
\email{dumin@yahoo.com, dumin@sai.msu.ru}

\author{Boris V. Somov}
\affil{P.K.~Sternberg Astronomical Institute (GAISh)
       of M.V.~Lomonosov Moscow State University,\\
       Universitetskii prosp.\ 13, 119992, Moscow, Russia}
\email{somov@sai.msu.ru}

\begin{abstract}
The probability of occurrence of various topological configurations of
the 3D null-point reconnection in a random magnetic field is studied.
It is found that the non-axisymmetrical six-tail configuration
(or ``improper radial null'')
should play the dominant role; while all other types of reconnection,
in particular, the axially-symmetric fan-like structures
(or ``proper radial nulls'') are realized with a much less probability.
A characteristic feature of the six-tail configuration is that at
the sufficiently large scales it is approximately reduced to
the well-known 2D X-type structure; and this explains why the 2D models
of reconnection usually work quite well.
\end{abstract}

\keywords{Magnetic fields --- Magnetic reconnection ---
          Sun: magnetic topology}

\section{INTRODUCTION}

It is commonly recognized that reconnection of the magnetic field lines
\citep[][and references therein]{pri00,som12,som13}
is of fundamental importance in the dynamics of various astrophysical
objects, ranging from planetary magnetospheres to interstellar medium,
as well as in the laboratory plasmas (\textit{e.g.},
\citealt{aul00,aul07};
\citealt{shi07};
\citealt{eyi13};
\citealt{zha12};
\citealt{wal03};
\citealt{dum02};
\citealt{erd07};
\citealt{ols13};
\citealt{mal13};
\citealt{ege12};
\citealt{liu13};
\citealt{gra14};
\citealt{osm14};
\citealt{hig13};
\citealt{lou13};
\citealt{mos12}).

A classical mechanism of the magnetic reconnection assumes its development
from the null (or ``neutral'')
point, where all components of the magnetic field~$ \bf B $ disappear.
(There are also some generalized models of reconnection which do not involve
the null points at all, \textit{e.g.}, as discussed by \citealt{pri03};
but we shall not consider such models in the present paper.)

Historically, the study of magnetic reconnection began from the 2D
approximation, where the null points possessed a universal topology of X-type.
However, starting from the mid 1990's a considerable attention was paid also
to the 3D case, where more diverse topological configurations are allowed
\citep[\textit{e.g.}, review by][]{pon11a}.

In the simplest case of a potential magnetic field, the structure of
field lines in the vicinity of 3D null point can be pictorially presented as
a collision of two oppositely-directed magnetic fluxes with subsequent
outflow in the equatorial plane.
This outflow (or ``fan'') can be either axially symmetric (which is called
the ``proper radial null'' according to terminology by \citealt{par96}) or
asymmetric (``improper radial null'').

It was implicitly assumed in many works that the most typical case of
the 3D null point, which can serve as a good initial approximation,
is just the axisymmetric fan-type structure (the proper radial null).
On the other hand, a few recent papers \citep{alh10,pon11b,gal11}
posed the problem of a ``generic'' 3D reconnection: they performed
a numerical simulation of the magnetic fields whose initial configurations
were substantially non-axisymmetric (\textit{i.e.}, represented
the improper radial nulls).
Unfortunately, it remained unclear how important are such configurations
from the statistical point of view?
In other words, how often do they appear in a random magnetic field?

It is the aim of the present paper to provide a self-consistent
calculation of the above-mentioned probabilities (and, thereby,
to give a justification for reasonable choice of the initial field
configurations in studies of the 3D magnetic reconnection).

\section{THEORETICAL ANALYSIS}

\subsection{The Previous Treatments}

A commonly-used approach to the analysis of the magnetic field structure
in the vicinity of a null point is its expansion in Taylor series
in the Cartesian coordinate system, whose origin is taken immediately
in the null point:
\begin{equation}
B_i = \sum\limits_j M_{ij} \, x_j ,
\;\; \mbox{where} \;\;
M_{ij} = \frac{\partial B_i}{\partial x_j} {\bigg|}_{{\bf x}=0}
\label{Eq:Taylor_expans}
\end{equation}
\citep[\textit{e.g.,}][and references therein]{gor88,par96}.
Since the magnetic field~{\bf B} must satisfy Maxwell equations,
elements of the matrix~{\bf M} are mutually related. In general,
this matrix can be described by four independent parameters;
see Eq.~(14) in the above-cited paper by \citet{par96}.

It can be intuitively assumed that the larger is the number of
the parameters involved and the greater are the domains of
their definition, the larger will be the probability of realization
of the respective field configuration.
However, it is not so easy to justify this conjecture by using
the representation~(\ref{Eq:Taylor_expans}):
since elements of the matrix~$ {\bf M} $ are mutually related by
Maxwell equations, it is not clear how one should choose their
joint probability distribution for the parameterization of the
random field.

To get around this obstacle, it is necessary to use an explicit solution
of the relevant field equations (\textit{e.g.}, in terms of the spherical
functions); so that the respective coefficients can be chosen as
independent random variables.
In the present paper, such an approach will be performed for the case of
potential (current-free) magnetic field.
A similar analysis for the non-potential field involves a more
cumbersome mathematics and, therefore, requires a separate paper.

\subsection{Initial Equations}

We shall consider \textit{random realizations} of the potential magnetic field
\begin{equation}
\mathbf{B} = - \mathrm{grad} \, \psi \, ,
\label{Eq:Def_Mag_field}
\end{equation}
where the magnetic potential~$ \psi $ satisfies the usual Laplace equation:
\begin{equation}
\Delta \psi = 0 \: .
\label{Eq:Laplace_eduation}
\end{equation}
(The potential field approximation is widely used, for example, in the solar
physics, although it may be less relevant for treating the magnetospheric
reconnection.)

Assuming the origin of spherical coordinate system~$ (r , \theta , \varphi) $
to be the spot of reconnection, solution of
equation~(\ref{Eq:Laplace_eduation}) can be written by the standard way as
\begin{equation}
{\psi} (r , \theta , \varphi) =
  \sum \limits_{j=0}^{\infty} \, \sum \limits_{m=0}^{j}
  r^j {\psi}_{jm} (\theta , \varphi) \: ,
\label{Eq:Potential_general}
\end{equation}
where
\begin{eqnarray}
&& \!\!\!\!\!\!\!\!\!\!\!\!\!\!\!\! {\psi}_{jm} (\theta , \varphi) =
\nonumber \\
&& \!\!\!\!\!\! P_j^m ( \cos \theta ) \,
  \big[ a_{jm} \cos ( m \varphi ) + b_{jm} \sin ( m \varphi ) \big]
\label{Eq:Ball_func_reduced}
\end{eqnarray}
are the spherical functions, and $ P_j^m $ are the adjoint Legendre
polynomials.
(The terms with negative powers of~$ r $ are not taken into account
because we are interested only in the nonsingular solutions.)
To avoid dealing with the infinite sum, it is convenient to assume that
expression~(\ref{Eq:Potential_general}) is cut off at a sufficiently large
value of~$ j $ and, therefore, contains only the finite number of terms~$ N $.
In other words, $ N $~is the dimensionality of the space of coefficients
$ a_{jm} $ and $ b_{jm} $.

If these coefficients are assumed to be random numbers, then we get a random
realization of the magnetic field~$ \mathbf{B} $.
It is a separate problem what are the reasonable probability distributions
for these coefficients.
However, it is important to emphasize that most of our subsequent results are
based only on the dimensionality of various subsets of the coefficients
$ a_{jm} $ and $ b_{jm} $, which are responsible for the various kinds of
reconnection.
Therefore, the respective conclusions should be valid for any nonsingular
probability distributions.

Let us begin to analyze the terms of magnetic
potential~(\ref{Eq:Potential_general}) with various powers of radius.
At $ j = 0 $, we get
$
{\psi}^{(0)} = a_{00} = \mathrm{const} ,
$
which evidently does not affect any physical results.

Next, at $ j = 1 $, the magnetic potential is
\begin{eqnarray}
{\psi}^{(1)} \!\!\!\!\!\! & = & \!\!\!\! r \, \big\{ a_{10} \cos \theta
\nonumber \\
  & - & \!\!\!\!\! ( 1 - {\cos}^2 \theta )^{1/2}
  [ a_{11} \cos \varphi + b_{11} \sin \varphi ] \big\} ;
\label{Eq:psi_1_init}
\end{eqnarray}
and its substitution into equation~(\ref{Eq:Def_Mag_field}) results in
\begin{eqnarray}
B_r^{(1)} \!\!\!\!\!\! & = & \!\!\! - \big\{ a_{10} \cos \theta
\nonumber \\
  & - & \!\!\!\!\! ( 1 - {\cos}^2 \theta )^{1/2}
  [ a_{11} \cos \varphi + b_{11} \sin \varphi ] \big\} .
\label{Eq:B_r_1_init}
\end{eqnarray}
Since $ r \! = 0 $ is assumed to be a null point (\textit{i.e.}, all
components of the magnetic field, including $ B_r $, should vanish),
we arrive at the requirement:
\begin{equation}
a_{10} = \, a_{11} = \, b_{11} = \, 0  \, .
\label{Eq:a1x_and_b1x}
\end{equation}
Because of these three constraints, \textit{the null point of any kind
will be realized only in a subspace of the random expansion coefficients
$ a_{jm} $ and $ b_{jm} $ with dimensionality $ N \! - 3 $ or less.}

At $ j = 2 $, the magnetic potential is written as
\begin{eqnarray}
{\psi}^{(2)} \!\!\!\!\! & = & \!\!\! r^2 \bigg\{
  \frac{1}{2} \, ( 3 \, {\cos}^2 \theta - 1 ) \, a_{20}
\nonumber \\
  & - & \!\!\! 3 \sin \theta \, \cos \theta \,
  \big[ a_{21} \cos \varphi
  + b_{21} \sin \varphi \big]
\nonumber \\
  & + & \!\!\! 3 \, {\sin}^2 \theta \,
  \big[ a_{22} \cos ( 2 \varphi ) + b_{22} \sin ( 2 \varphi ) \big]
\bigg\} .
\label{eq:psi_2_init}
\end{eqnarray}
Since we are interested in structure of the magnetic field lines rather
than in absolute values of the field, it is convenient to introduce
the normalized coefficients (denoted by a single subscript):
\begin{equation}
a_m \! = a_{2m} / a_{20} , \:\: b_m \! = b_{2m} / a_{20} , \quad
m = 1, 2 \, .
\label{eq:New_def_Coeff}
\end{equation}
Then, the magnetic field components take the form:
\begin{subequations}
\begin{eqnarray}
B_r^{(2)} \!\!\!\!\! & = & \!\!\!\!
  - 2 a_{20} r \, \Big\{ \frac{1}{2} \,
  ( 3 \, {\cos}^2 \theta - 1 )
\nonumber \\
  & - & \!\!\!\! \frac{3}{2} \, \sin ( 2 \theta ) \,
  \big[ a_1 \cos \varphi + b_1 \sin \varphi \big] +
\nonumber \\
  & + & \!\!\!\! 3 \: {\sin}^2 \theta \,
  \big[ a_2 \cos ( 2 \varphi ) + b_2 \sin ( 2 \varphi ) \big] \Big\} \: ,
\label{eq:B_r(2)}
\\
B_{\theta}^{(2)} \!\!\!\!\! & = & \!\!\!\!
  - 3 a_{20} r \, \Big\{ \sin ( 2 \theta )
  \big[ - \! \frac{1}{2} + a_2 \cos ( 2 \varphi )
\nonumber \\
  & + & \!\!\!\! b_2 \sin ( 2 \varphi ) \big]
\nonumber \\
  & - & \!\!\!\! \cos ( 2 \theta ) \big[ a_1 \cos \varphi
  + b_1 \sin \varphi \big] \Big\} \: ,
\label{eq:B_theta(2)}
\\
B_{\varphi}^{(2)} \!\!\!\!\! & = & \!\!\!\!
  - 3 a_{20} r \, \Big\{ 2 \, \sin \theta
  \big[ - \! a_2 \sin ( 2 \varphi ) + b_2 \cos ( 2 \varphi ) \big]
\nonumber \\
  & + & \!\!\!\! \cos \theta \big[ a_1 \sin \varphi
  - b_1 \cos \varphi \big] \Big\} \; .
\label{eq:B_phi(2)}
\end{eqnarray}
\end{subequations}

\subsection{Asymptotic Directions}

Following the standard procedures, the equation of a magnetic field
line can be written as
\begin{equation}
\frac{ d r }{B_r / ( a_{20} r )} \, = \,
\frac{ r \, d \theta }{B_{\theta} / ( a_{20} r )} \, = \,
\frac{ r \, \sin \theta \, d \varphi }{B_{\varphi} / ( a_{20} r )} \; .
\label{eq:field_eq_modif}
\end{equation}
Since quantities $ B^{(2)}_r \!\! / (a_{20} r)$,
$ B^{(2)}_{\theta} \! / (a_{20}r )$, and
$ B^{(2)}_{\varphi} \! / (a_{20}r )$ do not depend on $ r $,
in the limit $ r \! \to 0 $ we arrive at the conditions specifying
\textit{the field lines passing immediately through the null point}:
\begin{equation}
B^{(2)}_{\theta} \! / (a_{20} r) \, = \, 0 \: , \quad
B^{(2)}_{\varphi} \! / (a_{20}  r) \, = \, 0 \: .
\label{eq:field_eq_set}
\end{equation}
Substitution of the detailed expressions~(\ref{eq:B_theta(2)}) and
(\ref{eq:B_phi(2)}) into (\ref{eq:field_eq_set}) gives the following set
of algebraic equations:
\begin{subequations}
\begin{eqnarray}
&& \!\!\!\!\!\!\!\!\!\!\!\!\!\!\!
  \sin ( 2 {\theta}^* \! ) \, \big[ - \! \frac{1}{2} +
  \, a_2 \cos ( 2 {\varphi}^* \! ) + \, b_2 \sin ( 2 {\varphi}^* \! ) \big]
\nonumber \\
&& \!\!\!\!\!\!\!\!\!\!
  - \cos ( 2 {\theta}^* \! ) \, \big[ \,
  a_1 \cos {\varphi}^* + \, b_1 \sin {\varphi}^* \big] = 0 ,
\label{eq:field_eq_set_detailed_1}
\\[1ex]
&& \!\!\!\!\!\!\!\!\!\!\!\!\!\!\!
  2 \sin {\theta}^* \big[
  - \! a_2 \sin ( 2 {\varphi}^* \! ) + \, b_2 \cos ( 2 {\varphi}^* \! ) \big]
\nonumber \\
&& \!\!\!\!\!\!\!\!\!\!
  + \, \cos {\theta}^*
  \big[ a_1 \sin {\varphi}^* - \, b_1 \cos {\varphi}^* \big] = 0 ,
\label{eq:field_eq_set_detailed_2}
\end{eqnarray}
\end{subequations}
where $ {\theta}^* $ and $ {\varphi}^* $ are the angles at which
the field line enters (or leaves) the null point.

\begin{figure}[t]
\includegraphics[width=7.6cm]{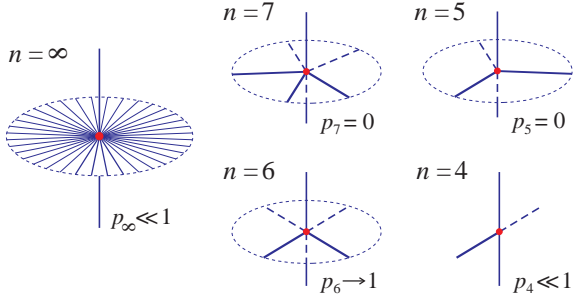}
\caption{
~Sketch of the various hypothetical null points, comprising both
the axially-symmetric fan-like configuration (left) and a few
structures with the finite number of asymptotic directions~$ n $
(right).
\label{fig:various_structures}}
\end{figure}

First of all, it can be easily checked that the above set of equations is
preserved under the transformation:
$ {\theta}^* \! \to \pi - {\theta}^*, \;
{\varphi}^* \! \to {\varphi}^* \! + \pi $.
Consequently, \textit{the magnetic field lines passing through the null
point always appear as the oppositely-directed pairs.}
So, the geometric structures with an odd number of tails (\textit{e.g.},
$ n = 5 $ or 7 in Figure~\ref{fig:various_structures}) cannot exist at all.

Next, let us analyze the particular solutions of
equations~(\ref{eq:field_eq_set_detailed_1}) and
(\ref{eq:field_eq_set_detailed_2}).
The simplest case evidently takes place at
$ a_1 = \, b_1 = \, a_2 = \, b_2 = \, 0 $ or, in the original designations,
\begin{equation}
a_{2m} \! = 0 , \;
b_{2m} \! = 0 , \:\: {\rm where}
\:\: m = 1, 2 \, .
\label{eq:Coeff_Fan_struct}
\end{equation}
Then, these equations are reduced to the simple condition
\begin{equation}
\sin ( 2 {\theta}^* \! ) = 0 \: ,
\label{eq:fan_struct}
\end{equation}
which has solutions of the two types:
\begin{equation}
{\theta}^* = 0 , \pi \quad
{\rm and} \quad
{\theta}^* = \pi / 2 \:\:
(\text{at any } {\varphi}^* ) \, .
\label{eq:Fan_struct_Solution}
\end{equation}
This represents a combination of the polar axis and a disk in
the equatorial plane, \textit{i.e.}, exactly \textit{the axially-symmetric
fan-like structure} depicted in the left-hand side of
Figure~\ref{fig:various_structures}.
(It is called ``the proper radial null'' according to terminology by
\citealt{par96}.)

Because of the 4 constraints~(\ref{eq:Coeff_Fan_struct}), this structure seems
to be realized in the subspace of coefficients $ a_{jm} $ and $ b_{jm} $ with
dimensionality $ N - 3 - 4 = N - 7 $.
However, it should be born in mind that these constraints were formulated for
the specific situation when ``spine'' of the ``fan'' was oriented exactly
along the polar axis of the coordinate system used.
In general, such fan-like structure can be rotated in space by two Euler
angles, which effectively removes 2 constraints.
So, \textit{the dimensionality of the relevant subset of coefficients will be
$ N - 7 + 2 = N - 5 $.}

Returning to the \textit{general case} of arbitrary coefficients $ a_1 $,
$ b_1 $, $ a_2 $, and $ b_2 $, it can be naturally assumed that the set of
two algebraic equations~(\ref{eq:field_eq_set_detailed_1}) and
(\ref{eq:field_eq_set_detailed_2}) for two unknown variables $ {\theta}^* $
and $ {\varphi}^* $ should have a finite number of solutions (\textit{i.e.},
the number of asymptotic tails in Figure~\ref{fig:various_structures} should
be finite).
Moreover, as follows from a more careful mathematical analysis, this number
is always equal to~6 (except for some special subset of coefficients $ a_i $
and $ b_i $ with lower dimensionality).

To prove this fact, it is convenient to reduce the above-mentioned system of
equations to a single equation for the azimuthal angle~$ {\varphi}^* $:
\begin{equation}
F( \eta ( {\varphi}^* ) , \zeta ( {\varphi}^* ) ) = 0 \, ,
\label{eq:Gen_eq_for_phi}
\end{equation}
where $ \eta = \cos {\varphi}^* $, $ \zeta \! = \sin {\varphi}^* $, and
\begin{eqnarray}
\!\!\!\!\!\!\!\! && F( \eta , \zeta ) =
  4 \big[ 2 a_2 \eta \zeta - b_2 ({\eta}^2 - {\zeta}^2) \big]
  ( a_1 \zeta - b_1 \eta )
\nonumber \\
\!\!\!\!\!\!\!\! && \quad \times \big[ \! - \! \frac{1}{2} +
  a_2 ({\eta}^2 - {\zeta}^2) + 2 b_2 \eta \zeta \big]
\nonumber \\
\!\!\!\!\!\!\!\! && \quad - \Big\{ 4 \big[ 2 a_2 \eta \zeta -
  b_2 ({\eta}^2 - {\zeta}^2) \big]^2 - ( a_1 \zeta - b_1 \eta )^2 \Big\}
\nonumber \\
\!\!\!\!\!\!\!\! && \quad \times ( a_1 \eta + b_1 \zeta ) \, .
\label{eq:Def_fun_F}
\end{eqnarray}
Then, if the roots~$ {\varphi}^* $ have been found, the corresponding values
of the polar angle~$ {\theta}^* $ can be easily restored from one of
equations~(\ref{eq:field_eq_set_detailed_1}) or
(\ref{eq:field_eq_set_detailed_2}).

\begin{figure}[t]
\includegraphics[width=7.6cm]{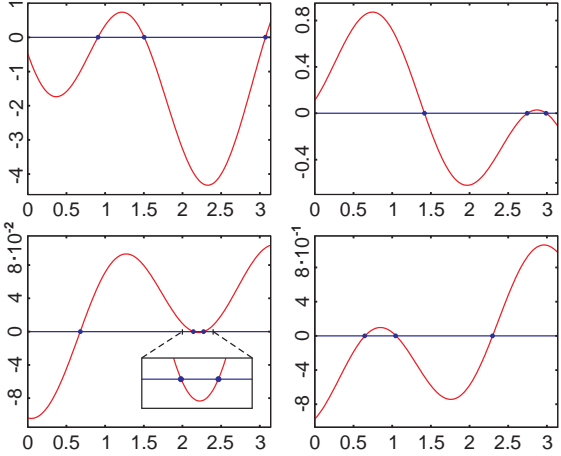}
\caption{A few examples of the function
$ F( {\eta}({\varphi}^*), $ ${\zeta}({\varphi}^*) ) $ at the
interval~$ {\varphi}^* \! \in [0, {\pi}] $ at random values of the
coefficients~$ a_1 $, $ a_2 $, $ b_1 $, and $ b_2 $.
\label{fig:function_F}}
\end{figure}

Since formula~(\ref{eq:Def_fun_F}) represents a quite complex polynomial
expression, the simplest approach to resolve our task is just to perform
a statistical simulation: let us generate a sufficiently large sequence
of random coefficients $ a_1 $, $ a_2 $, $ b_1 $, and $ b_2 $ (\textit{e.g.},
as a Gaussian distribution with a zero mean) and then analyze behavior of
the function~$ F( {\eta}({\varphi}^*), {\zeta}({\varphi}^*) ) $ graphically
(Figure~\ref{fig:function_F}).
Surprisingly, it was found that the plot of~$ F $ intersects the horizontal
axis always in 3~points at the interval~$ {\varphi}^* \! \in [0, {\pi}] $
(and, consequently, in 6~points at the
interval~$ {\varphi}^* \! \in [0, 2{\pi}] $).
In fact, a subsequent careful analysis enabled us to get a rigorous
mathematical proof of this fact.
However, because of the cumbersome formulas, it will be not presented
here, and we prefer to appeal just to the results of statistical simulation.
Besides, it was established that the above-mentioned six solutions of
the equations~(\ref{eq:field_eq_set_detailed_1}) and
(\ref{eq:field_eq_set_detailed_2}) correspond geometrically to the six
``tails'' which are mutually orthogonal to each other.

Therefore, we have found that \textit{a generic 3D null point}
in the potential field approximation should have a specific six-tail structure
(\textit{i.e.}, possess 6 asymptotic directions of the magnetic field).
This is because it is realized in the subspace of coefficients of the random
field with dimensionality $ N \! - 3 $, \textit{i.e.}, almost in the entire
space allowed for the null point by the constraints~(\ref{Eq:a1x_and_b1x}).
All other configurations (in particular, the intuitively attractive
axially-symmetric fan or more exotic geometric structures outlined in the old
paper by~\citealt{zhu66}) should emerge with a much less probability, because
they are realized in the subsets of coefficients with lower dimensionality.
It is especially important to emphasize that, since these conclusions are
based only on the dimensionality of the relevant subspaces, they should be
valid for any nonsingular probability distribution of the random-field
coefficients.
(So, the particular Gaussian distribution used in the simulation presented
in Figure~\ref{fig:function_F} does not affect the final result.)

\subsection{Structure of the Field Lines}

It is important, of course, to discuss a pattern of the magnetic field
lines \textit{in the vicinity} of the above-mentioned generic configuration.
To avoid cumbersome formulas, let us consider the simplest (but completely
representative) case $ a_1 \! = b_1 \! = b_2 \! = 0, \: a_2 \! \neq 0 $,
which corresponds to the six-tail structure oriented along the axes of the
coordinate system.
Then, equations~(\ref{eq:field_eq_set_detailed_1}) and
(\ref{eq:field_eq_set_detailed_2}) are simplified to
\begin{subequations}
\begin{eqnarray}
&& \!\!\!\!\!\!\!\!\!\!
  \sin ( 2 {\theta}^* \! ) \, \big[ - \! \frac{1}{2} +
  \, a_2 \cos ( 2 {\varphi}^* \! ) \, \big] = 0 \, ,
\label{eq:field_eq_set_simplified_1}
\\[0.5ex]
&& \!\!\!\!\!\!\!\!\!\!
  a_2 \, \sin {\theta}^* \sin ( 2 {\varphi}^* \! ) = 0 \, .
\label{eq:field_eq_set_simplified_2}
\end{eqnarray}
\end{subequations}
Their solutions are evidently $ {\theta}^* \! = 0, \pi $ and
$ {\theta}^* \! = \pi \! / 2 , \:
{\varphi}^* \! = 0, \pi \! / 2 , \pi , 3 \pi \! / 2 $,
which correspond just to the six semiaxes of the coordinate system.

Next, omitting the unessential common multiplier~$ a_{20} $,
expressions~(\ref{eq:B_r(2)})--(\ref{eq:B_phi(2)}) for the magnetic field
components are reduced to
\begin{subequations}
\begin{eqnarray}
&& \!\!\!\!\!\!\!\!\!\!\!
B_r \! = - 2 r \big[ \, \frac{1}{2} \, ( 3 \, {\cos}^2 \theta - 1 )
\nonumber \\
&& \;\:
+ \, 3 a_2 \, {\sin}^2 \theta \, \cos ( 2 \varphi ) \big] \, ,
\label{eq:B_r(2)_simpl}
\\
&& \!\!\!\!\!\!\!\!\!\!\!
B_{\theta} \! = - 3 r \sin ( 2 \theta ) \big[ - \! \frac{1}{2}
  + a_2 \cos ( 2 \varphi ) \big] \, ,
\label{eq:B_theta(2)_simpl}
\\[0.5ex]
&& \!\!\!\!\!\!\!\!\!\!\!
B_{\varphi} \! = \, 6 a_2 r \sin \theta \, \sin ( 2 \varphi ) \, .
\label{eq:B_phi(2)_simpl}
\end{eqnarray}
\end{subequations}
As expected, $ B_{\theta} $ and $ B_{\varphi} $ vanish immediately at the
coordinate axes, while $ B_r $ changes its sign on the opposite sides from
the origin.

\begin{figure}[t]
\begin{center}
\includegraphics[width=6.8cm]{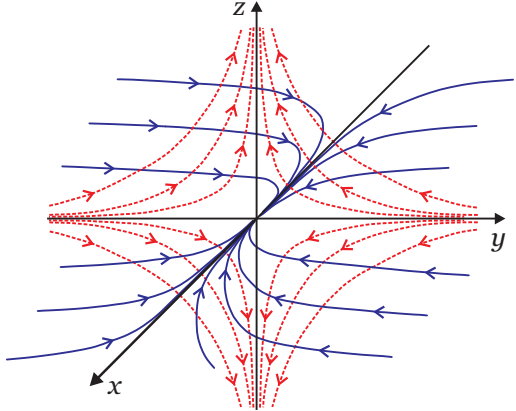}
\end{center}
\caption{
Sketch of magnetic field lines in the six-tail configuration.
Solid (blue) curves represent the field lines in the horizontal
$ x y $-plane; and dotted (red) curves, in the vertical $ y z $-plane.
Field lines in another vertical plane~$ x z $, perpendicular to the
plane of figure, are not shown here; they have the same hyperbolic
structure as in the $ y z $-plane.
\label{fig:field_line_struct}}
\end{figure}

Substituting expressions~(\ref{eq:B_r(2)_simpl})--(\ref{eq:B_phi(2)_simpl})
into~(\ref{eq:field_eq_modif}) and performing the integration, we can easily
find formulas for the magnetic field lines in three coordinate planes.
For example, in the $ xy $-plane ($ \theta \! = \pi \! / 2 $) the final
result will take the form:
\begin{equation}
r = C \Big( \big| \sin \varphi \big| ^{1 - 1 / ( 6 a_2 ) } \,
    \big| \cos \varphi \big| ^{1 + 1 / ( 6 a_2 )} \Big) ^{\! -1/2} ,
\label{eq:field_line_xy_plane}
\end{equation}
where $ C $~is an arbitrary constant.
Behaviour of this function has three qualitatively different regimes,
depending on the value of coefficient~$ a_2 $:
\begin{itemize}
\item[(a)]
If $ a_2 \! < \! -1/6 $ or $ a_2 \! > \! 1/6 $, then
$ r \! \rightarrow \infty $ both at $ \varphi \! \rightarrow 0 $ and
$ \varphi \! \rightarrow \pi / 2 $.
This evidently corresponds to the field line of hyperbolic type.
\item[(b)]
If $ -1/6 < a_2 \! < 0 $, then $ r \! \rightarrow \infty $ at
$ \varphi \! \rightarrow 0 $ and $ r \! \rightarrow 0 $ at
$ \varphi \! \rightarrow \pi / 2 $.
This is the field line of parabolic type with the parabola axis oriented
in $ x $-direction.
\item[(c)]
If $ 0 < a_2 \! < \! 1/6 $, then $ r \! \rightarrow 0 $ at
$ \varphi \! \rightarrow 0 $ and $ r \! \rightarrow \infty $ at
$ \varphi \! \rightarrow \pi / 2 $.
This is also the field line of parabolic type but with the parabola axis
oriented in $ y $-direction.
\end{itemize}
As regards the field lines in two other coordinate planes, they can be shown
to have a hyperbolic structure in cases~(b) and (c).
Just this situation is illustrated in Figure~\ref{fig:field_line_struct}.

Furthermore, it can be proved that the same pattern of the field lines can be
associated with all other cases mentioned above just by interchanging the role
of various coordinate axes.
Using the terminology adopted in the theory of differential equations, one
can say that the field lines have a node structure in one of the coordinate
planes and the saddle structure in two other planes.

Let us emphasize that the six-tail arrangement of the magnetic field lines
in the vicinity of a 3D null point is not a new finding:
this configuration is well known from the earlier works, \textit{e.g.},
Figure~1 in paper by \citet{gor88} or Figure~5 in paper by \citet{par96}),
where it was called ``the improper radial null''.
However, it has not been recognized up to now that just this structure
should play the dominant role in the 3D magnetic reconnection.

Besides, the previously-used term ``improper'' looks somewhat misleading
in the case that is actually the most typical.
So, we prefer to call it ``the six-tail configuration''.
In fact, the purely geometric aspects of this structure may be studied
more efficiently by the employment of Taylor expansion in Cartesian
coordinates~(\ref{Eq:Taylor_expans}); for more details, see Section~III
in the above-cited paper by \citet{par96}.
On the other hand, our approach, based on the spherical functions,
makes it possible to perform an accurate statistical parameterization of
the random magnetic field and, thereby, to calculate the respective
probabilities.

\subsection{Pictorial Illustration}

\begin{figure}[t]
\includegraphics[width=7.6cm]{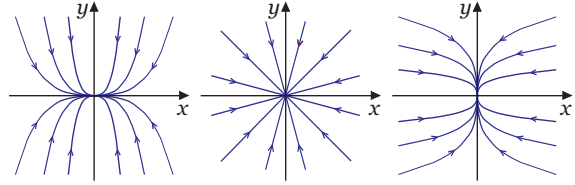}
\caption{Appearance of the axially-symmetric fan as the intermediate
case between two six-tail configurations.
Only field lines in the $ x y $-plane are drawn here; the field lines
in the $ x z $- and $ y z $-planes always remain of the saddle type.
\label{fig:struct_transition}}
\end{figure}

It can be easily understood why probability of occurrence of
the axially-symmetric fan-type structure (left-hand panel in
Fig.~\ref{fig:various_structures}) should be substantially suppressed
as compared to the six-tail structure.
Really, let us pay attention to the behavior of field lines in the
$ xy $-plane of Figure~\ref{fig:field_line_struct}
and assume that initially the value of parameter~$ a_2 $
corresponded to the case~(c), as depicted in the left-hand panel of
Figure~\ref{fig:struct_transition}.
Next, let $ a_2 $ gradually decrease and become negative, which refers
to the case~(b).
From the geometric point of view, this corresponds to a gradual decrease in
curvature of the field lines, and at some instant they become bent in another
direction, \textit{i.e.}, the entire pattern remains parabolic but the
parabola axis jumps by~$ \pi / 2 $ (right-hand panel in
Figure~\ref{fig:struct_transition}).
Then, the boundary between these two cases is just the axially-symmetric
fan-type structure, depicted in the central panel, which is realized at
$ a_2 \! = 0 $.

In other words, there are infinitely many six-tail configurations of
types~(b) and~(c) but only one intermediate axially-symmetric fan-type
configuration.
So, the probability of its realization should be extremely small.
(This picture is quite similar to Figure~2 by~\citealt{pri96}, who
used the magnetic field parameterization in Cartesian coordinates.)

\subsection{Reduction to Quasi-2D Geometry}

Returning to Figure~\ref{fig:field_line_struct}, attention should be paid to
the fact that six asymptotic directions of the magnetic field are quite
different from each other.
Namely, four of them ($ y $, $ -y $, $ z $, and $ -z $) can be called
``dominant'', because most of the field lines tend to approach one of these
directions when they go away from the null point.
On the other hand, two other asymptotic directions ($ x $ and $ -x $) should
be called ``recessive'', because most of the field lines tend to depart from
them.
Therefore, the recessive directions will be ``lost'' when observed from
a large distance, and the entire pattern will look like a classical
2D X-point.
This fact can explain why the 2D models of magnetic reconnection usually
work rather well.

For example, \cite{mas09} performed a numerical simulation of the solar
flare presumably caused by a single null point and have found that
the respective 3D reconnection actually proceeds in
the quasi-two-dimensional slabs.
The same reduction to a quasi-2D configuration was observed very clearly
in our numerical simulations of bifurcation of the 3D null points in
the solar atmosphere \citetext{Dumin \& Somov 2015, in preparation}.

\section{CONCLUSIONS}

We have calculated the probability of occurrence of various kinds of the 3D
null points in a random magnetic field and studied the structure
of field lines in their vicinity.
As a result, it was found that:
\begin{enumerate}
\item
Contrary to the intuitive expectations, the most likely case of the 3D null
point, responsible for the magnetic reconnection, is the specific six-tail
structure (or ``the improper radial null'', according to terminology by
\citealt{par96}).
\item
All other kinds of the 3D null points, in particular, the intuitively
attractive axially-symmetric fan (or ``proper radial null'')
are realized with much less probabilities, as schematically summarized
in Figure~\ref{fig:various_structures}.
\item
At the sufficiently large distances, the generic six-tail structure is
approximately reduced to a quasi-2D configuration of the well-known X-type,
which explains why the 2D approach is often a good approximation for the
magnetic reconnection.
Therefore, it may be conjectured that the specific 3D effects should be
important, first of all, in the small-scale magnetic reconnection events
(\textit{e.g.}, in the solar micro- and nano-flares).
\end{enumerate}

\acknowledgments

One of the authors (YVD) is grateful to
J.~B{\"u}chner,
A.T.~Lukashenko,
D.D.~Sokoloff, and
I.S.~Veselovsky
for fruitful discussions and comments.
We are also grateful to the unknown referee for valuable bibliographic
suggestions.

\end{document}